\documentclass{article}

\begin{document}
\title{Signals of the Abelian $Z'$ boson within the analysis of the LEP2 data}
\author{V.I. Demchik, A.V. Gulov\thanks{Email: gulov@ff.dsu.dp.ua},
 V.V. Skalozub\thanks{Email: skalozub@ff.dsu.dp.ua}, and A.Yu. Tischenko\\
{\it Dniepropetrovsk National University,}\\ {\it Dniepropetrovsk,
49050 Ukraine}}
\date{February 24, 2003}
\maketitle
\begin{abstract}
The preliminary LEP data on the $e^+e^-\rightarrow l^+l^-$
scattering are analysed to establish a model-independent search
for the signals of virtual states of the Abelian $Z'$ boson. The
recently introduced observables give a possibility to pick up
uniquely the Abelian $Z'$ signals in these processes. The mean
values of the observables are in accordance with the $Z'$
existence. However, the accuracy of the experimental data is
deficient to detect the signal at more than the $1\sigma$
confidence level. The results of other model-independent fits and
further prospects are discussed.
\end{abstract}

\section{Introduction}

The recently stopped LEP2 experiments have accumulated a huge
amount of data on four-fermion processes at the center-of-mass
energies $\sqrt{s}\sim 130-207$ GeV \cite{EWWG,LEP2FF02-03}.
Besides the precision tests of the Standard Model (SM) of
elementary particles these data allow the estimation of the energy
scale of a new physics beyond the SM.

Various approaches to detect manifestations of physics beyond the
SM have been proposed in the literature. They could be subdivided
into model-dependent and model-independent methods. The former
usually means the comparison of experimental data with the
predictions of some specific models which extend the SM at high
energies. In this way some popular Grand Unified theories, the
supersymmetry models as well as the theories of technicolor or
extra dimensions are intensively discussed and the values of
couplings, mixing angles, and particle masses are constrained. In
particular, model-dependent bounds are widely presented in the LEP
reports \cite{EWWG,LEP2FF02-03}.

In the model-independent approach one fits some effective
low-energy parameters such as four-fermion contact couplings.
Below the scale of the heavy particle decoupling various theories
beyond the SM can be described by the same set of effective
contact interactions between the SM particles. The only difference
is in the values of the corresponding coefficients, which can be
fitted by experimental data. This idea is elaborated in the
Effective Lagrangian method \cite{EL} as well as in the helicity
`models' introduced by the LEP Collaborations (LL, RR, \ldots)
\cite{EWWG,LEP2FF02-03}. An advantage of the approach is the
restricted set of parameters which describe the low-energy
phenomenology of any model beyond the SM for a specified
scattering process. Unfortunately, each effective
model-independent parameter conceals a number of different
scenarios of new physics. As a consequence, the model-independent
approach gives a possibility to detect a signal of new physics,
but it cannot distinguish the particle (defined by specific
quantum numbers) responsible for the signal.

It seems to us that it is reasonable to develop the
model-independent searches for the manifestations of heavy
particles with specific quantum numbers. Such an approach is
intended to detect the signal of some heavy particle (for example,
the heavy neutral vector boson) by means of the data of the LEP2
or other experiments without specifying a model beyond the SM. In
this way, it is also possible to derive model-independent
constraints on the mass and the couplings of the considered heavy
particle. To develop this approach one has to take into account
some model-independent relations between the couplings of the
heavy particle as well as some features of the kinematics of the
considered scattering processes.

In the present paper we focus on the problem of model-independent
searches for signals of the heavy Abelian $Z'$ boson \cite{leike}
by means of the analysis of the LEP2 data on the lepton processes
$e^+ e^- \rightarrow \mu^+\mu^-$ and $e^+ e^- \rightarrow
\tau^+\tau^-$. This particle is a necessary element of different
models extending the SM. The low limits on its mass estimated for
a variety of popular models ($\chi$, $\psi$, $\eta$, L--R models
\cite{models} and the Sequential Standard Model (SSM) \cite{SSM})
are found to be in the wide energy interval 600--2000 GeV
\cite{EWWG,LEP2FF02-03}. In what follows we assume that the $Z'$
boson is heavy enough to be decoupled at the LEP2 energies.

In the previous papers \cite{ZprTHDM} we argued that the
low-energy $Z'$ couplings to the SM particles satisfy some
model-independent relations, which are the consequences of
renormalizability of a theory beyond the SM remaining in other
respects unspecified. These relations, called the renormalization
group (RG) relations, predict two possible types of the low-energy
$Z'$ interactions with the SM fields, namely, the chiral and the
Abelian $Z'$ bosons. Each $Z'$ type is described by a few
couplings to the SM fields. Therefore, it is possible to introduce
observables which uniquely pick up the $Z'$ virtual state
\cite{ZprTHDM}.

The $Z'$ signal in the four-lepton scattering process $e^+e^-\to
l^+l^-$ can be detected with a sign-definite observable, which is
ruled by the center-of-mass energy and an additional kinematic
parameter. The incorporation of the next-to-leading terms in
$m^{-2}_{Z'}$ allows to consider the $Z'$ effects beyond the
approach of four-fermion contact interactions \cite{rizzo}. As a
consequence, the four-fermion contact couplings and the $Z'$ mass
can be fitted separately.

The introduced observable can be computed directly from the
differential cross-sections. However, the statistical errors of
the available differential cross-sections of LEP2 experiments are
of one order larger than the accuracy of the corresponding total
cross-sections and the forward-backward asymmetries. Fortunately,
the differential cross-sections of the $e^+e^-\to l^+l^-$
processes at the LEP2 energies (including the one-loop radiative
corrections) can be successfully approximated by a two-parametric
polynomial of the cosine of the scattering angle. It gives a
possibility to recalculate the observable from the total
cross-sections and the forward-backward asymmetries reducing
noticeably the experimental uncertainty.

Thus, the outlined analysis has to answer whether or not one could
detect the model-independent signal of the Abelian $Z'$ boson by
treating the LEP2 data. As it will be shown, the LEP2 data on the
scattering into $\mu$ and $\tau$ pairs lead to the Abelian $Z'$
signal at about 1$\sigma$ confidence level.

The paper is organized as follows. In sect. 2 the necessary
information on the model-independent description of the $Z'$
interactions at low energies and the RG relations are given. In
sect. 3 the observables to pick up uniquely the $Z'$ boson are
introduced. In the last section the results on the LEP data fit
and the conclusions as well as further prospects are discussed.

\section{$Z'$ couplings to the SM particles}

The Abelian $Z'$ boson can be introduced in a model-independent
(phenomenological) way by defining its effective low-energy
couplings to the SM particles. Such a parameterization is well
known in the literature \cite{leike}. Since we are going to
account of the $Z'$ effects in the $e^+e^-\to l^+l^-$ process at
LEP2 energies $\sqrt{s}\ll m_{Z'}$, we parameterize the $Z'$
interactions induced at the tree-level, only. As the decoupling
theorem \cite{decoupling} guarantees, they are of renormalizable
type, since the non-renormalizable interactions are generated at
higher energies due to radiation corrections and suppressed by the
inverse heavy mass ($m_{Z'}$ in our case). The SM gauge group
$SU(2)_L\times U(1)_Y$ is considered as a subgroup of the
underlying theory group. So, the mixing interactions of the types
$Z'W^+W^-$, $Z'ZZ$, ... are absent at the tree level. Under these
assumptions the $Z'$ couplings to the fermion and scalar fields
are described by the Lagrangian:
\begin{eqnarray}\label{1}
 {\cal L}&=& \left|\left( D^{{\rm ew,} \phi}_\mu -
  \frac{i\tilde{g}}{2}\tilde{Y}(\phi)\tilde{B}_\mu
  \right)\phi\right|^2 + \nonumber\\&&
 i\sum\limits_{f=f_L,f_R}\bar{f}{\gamma^\mu}
  \left(
  D^{{\rm ew,} f}_\mu -
  \frac{i\tilde{g}}{2}\tilde{Y}(f)\tilde{B}_\mu
  \right)f,
\end{eqnarray}
where $\phi$ is the SM scalar doublet, $\tilde{B}_\mu$ denotes the
massive $Z'$ field before the spontaneous breaking of the
electroweak symmetry, and the summation over the all SM
left-handed fermion doublets, $f_L =\{(f_u)_L, (f_d)_L\}$, and the
right-handed singlets, $f_R = (f_u)_R, (f_d)_R$, is understood.
The notation $\tilde{g}$ stands for the charge corresponding to
the $Z'$ gauge group, $D^{{\rm ew,}\phi}_\mu$ and $D^{{\rm
ew,}f}_\mu$ are the electroweak covariant derivatives. Diagonal
$2\times 2$ matrices $\tilde{Y}(\phi)={\rm
diag}(\tilde{Y}_{\phi,1},\tilde{Y}_{\phi,2})$,
$\tilde{Y}(f_L)={\rm diag}(\tilde{Y}_{L,f_u},\tilde{Y}_{L,f_d})$
and numbers $\tilde{Y}(f_R)=\tilde{Y}_{R,f}$ mean the unknown $Z'$
generators characterizing the model beyond the SM.

In particular, the Lagrangian (\ref{1}) generally leads to the
$Z$--$Z'$ mixing of order $m^2_Z/m^2_{Z'}$ which is proportional
to $\tilde{Y}_{\phi,2}$ and originated from the diagonalization of
the neutral vector boson states. The mixing contributes to the
scattering amplitudes and cannot be neglected at the LEP2 energies
\cite{NPB}.

The $Z'$ couplings to a fermion $f$ are parameterized by two
numbers $\tilde{Y}_{L,f}$ and $\tilde{Y}_{R,f}$. Alternatively,
the couplings to the axial-vector and vector fermion currents,
$a^l_{Z'}\equiv(\tilde{Y}_{R,l}-\tilde{Y}_{L,l})/2$ and
$v^l_{Z'}\equiv(\tilde{Y}_{L,l}+\tilde{Y}_{R,l})/2$, can be used.
Their values are determined by the unknown model beyond the SM.
Assuming an arbitrary underlying theory one usually supposes that
the parameters $a_f$ and $v_f$ are independent numbers. However,
if a theory beyond the SM is renormalizable these parameters
satisfy some relations. For $Z'$ boson this is reflected in the
correlations between $a_f$ and $v_f$ \cite{ZprTHDM}. These
correlations are model-independent in a sense that they do not
depend on the underlying model. The detailed discussion of this
point and the derivation of the RG relations are given in Ref.
\cite{ZprTHDM}. Therein it is shown that two types of the $Z'$
boson interactions are admitted -- the chiral and the Abelian
ones. In the present paper we are interested in the Abelian $Z'$
couplings which are described by the relations:
\begin{equation}\label{2}
v_f-a_f=v_{f^\star}-a_{f^\star},\quad a_f=T_{3,f}\tilde{Y}_\phi,
\quad \tilde{Y}_{\phi,1}=\tilde{Y}_{\phi,2}\equiv\tilde{Y}_\phi,
\end{equation}
where $T^3_f$ is the third component of the fermion weak isospin,
and $f^\star$ means the isopartner of $f$ (namely,
$l^\star=\nu_l,\nu^\star_l=l,\ldots$).

The relations (\ref{2}) ensure, in particular, the invariance of
the Yukawa terms with respect to the effective low-energy
$\tilde{U}(1)$ subgroup corresponding to the Abelian $Z'$ boson.
As it follows from the relations, the couplings of the Abelian
$Z'$ to the axial-vector fermion currents have the universal
absolute value proportional to the $Z'$ coupling to the scalar
doublet. So, in what follows we will use the short notation
$a=a_l=-\tilde{Y}_\phi/2$. Note also that the $Z$--$Z'$ mixing is
expressed in terms of the axial-vector coupling $a$.

An important benefit of the relations (\ref{2}) is the possibility
to reduce the number of independent parameters of new physics. For
example, they can be used to relate the coefficients of the
effective Lagrangians \cite{ijmpa}. Due to a fewer number of
independent $Z'$ couplings the amplitudes and cross-sections of
different scattering processes are also related. As a result, one
is able to pick up the characteristic signal of the Abelian
$Z'$-boson in these processes and to fit successfully the
corresponding $Z'$ couplings. In the present paper we take into
account the RG relations (\ref{2}) in order to introduce the
observables convenient for detecting uniquely the Abelian $Z'$
signals in LEP experiments and to obtain the corresponding
experimental constraints on the signal.

\section{Observables}

\subsection{The differential cross section}

Let us consider the processes $e^+e^-\to l^+l^-$ ($l=\mu,\tau$)
with the non-polarized initial- and final-state fermions. In order
to introduce the observable which selects the signal of the
Abelian $Z'$ boson we need to compute the differential
cross-sections of the processes up to the one-loop level.

The lower-order diagrams describe the neutral vector boson
exchange in the $s$-channel ($e^+e^-\to V^\ast\to l^+l^-$,
$V=A,Z,Z'$). As for the one-loop corrections, two classes of
diagrams are taken into account. The first one includes the pure
SM graphs (the mass operators, the vertex corrections, and the
boxes). The second set of the one-loop diagrams improves the
Born-level $Z'$-exchange amplitude by ``dressing'' the $Z'$
propagator and and the $Z'$--fermion vertices. We assume that $Z'$
states are not excited inside loops. Such an approximation means
that the $Z'$-boson is completely decoupled. Then, the
differential cross-section consists of the squared tree-level
amplitude and the term from the interference of the tree-level and
the one-loop amplitudes. To obtain an infrared-finite result, we
also take into account the processes with the soft-photon emission
in the initial and final states.

Various computational software for calculation of amplitudes and
cross-sections has been developed. For example, the SM
cross-sections in the LEP fits are usually generated with ZFITTER.
However, ZFITTER requires severe modifications to incorporate the
effects of heavy particles beyond the SM. Therefore, we perform
the necessary calculations with other software. The Feynman
diagrams and the amplitudes are generated with FEYNARTS. The
algebraic reduction of the one-loop tensor integrals to the scalar
integrals as well as the cross-section construction are carried
out with FORMCALC. The scalar one-loop integrals are evaluated
with LOOPTOOLS library within the $\overline{\mbox{MS}}$
renormalization scheme. The unknown Higgs boson mass is set to
125GeV in accordance with the present day bounds.

In the lower order in $m^{-2}_{Z'}$ the $Z'$ contributions to the
differential cross-section of the process $e^+e^-\to l^+l^-$ are
expressed in terms of four-fermion contact couplings, only. If one
takes into consideration the higher-order corrections in
$m^{-2}_{Z'}$, it becomes possible to estimate separately the
$Z'$-induced contact couplings and the $Z'$ mass \cite{rizzo}. In
the present analysis we keep the terms of order $O(m^{-4}_{Z'})$
to fit both of these parameters.

Expanding the differential cross-section in the inverse $Z'$ mass
and neglecting the terms of order $O(m^{-6}_{Z'})$, we have
\begin{eqnarray}
\frac{d\sigma_l(s)}{dz} &=& \frac{d\sigma_l^{\rm SM}(s)}{dz}
+\sum_{i=1}^{7}\sum_{j=1}^{i}
\left[A_{ij}^l(s,z)+B_{ij}^l(s,z)\zeta\right]a_{i}a_{j}
\nonumber\\&&
+\sum_{i=1}^{7}\sum_{j=1}^{i}\sum_{k=1}^{j}\sum_{n=1}^{k}
C_{ijkn}^l(s,z)a_{i}a_{j}a_{k}a_{n},
\end{eqnarray}
where the dimensionless quantities
\begin{eqnarray}
 \zeta &=& \frac{m^2_Z}{m^2_{Z'}},\quad
 \epsilon=\frac{\tilde{g}^2 m^2_Z a^2}{4\pi m^2_{Z'}},
\nonumber\\
 (a_1,a_2,a_3,a_4,a_5,a_6,a_7) &=&
 \sqrt\frac{\tilde{g}^2 m^2_Z}{4\pi m^2_{Z'}}
 (a,v_e,v_\mu,v_\tau,v_d,v_s,v_b)
\end{eqnarray}
are introduced. In what follows the index $l=\mu,\tau$ denotes the
final-state lepton.

The coefficients $A$, $B$, $C$ are determined by the SM couplings
and masses. Each factor may include the tree-level contribution,
the one-loop correction and the term describing the soft-photon
emission. The factors $A$ describe the leading-order contribution,
whereas others correspond to the higher order corrections in
$m^{-2}_{Z'}$.

\subsection{The observable}

To take into consideration the correlations (\ref{2}) we introduce
the observable $\sigma_l(z)$ defined as the difference of cross
sections integrated in some ranges of the scattering angle
$\theta$ \cite{obs}:
\begin{eqnarray}\label{eq8}
 \sigma_l(z)
 &\equiv&\int\nolimits_z^1
  \frac{d\sigma_l}{d\cos\theta}d\cos\theta
 -\int\nolimits_{-1}^z
  \frac{d\sigma_l}{d\cos\theta}d\cos\theta,
\end{eqnarray}
where $z$ stands for the cosine of the boundary angle. The idea of
introducing the $z$-dependent observable (\ref{eq8}) is to choose
the value of the kinematic parameter $z$ in such a way that to
pick up the characteristic features of the Abelian $Z'$ signals.

The deviation of the observable from its SM value can be derived
by the angular integration of the differential cross-section and
has the form:
\begin{eqnarray}
\Delta\sigma_l(z) &=& \sigma_l(z) - \sigma^{\rm SM}_l(z)
=\sum_{i=1}^{7}\sum_{j=1}^{i}
\left[\tilde{A}_{ij}^l(s,z)+\tilde{B}_{ij}^l(s,z)\zeta\right]a_{i}a_{j}
\nonumber\\&&
+\sum_{i=1}^{7}\sum_{j=1}^{i}\sum_{k=1}^{j}\sum_{n=1}^{k}
\tilde{C}_{ijkn}^l(s,z)a_{i}a_{j}a_{k}a_{n}.
\end{eqnarray}

There is an interval of values of the boundary angle, at which the
factors $\tilde{A}^l_{11}$, $\tilde{B}^l_{11}$, and
$\tilde{C}^l_{1111}$ at the sign-definite parameters $\epsilon$,
$\epsilon\zeta$, and $\epsilon^2$ contribute more than 95\% of the
observable value. It gives a possibility to construct the
sign-definite observable $\Delta\sigma_l(z^*)<0$ by specifying the
proper value of $z^*$.

In general, one could choose the boundary angle $z*$ in different
schemes. In the previous papers \cite{obs,NPB} we considered just
a few number of tree-level four-fermion contact couplings and
specified $z^*$ in order to cancel the factor at the vector-vector
coupling. However, if one-loop corrections are taken into account
there are a large number of additional contact couplings. So, we
have to define some quantitative criterion $F(z)$ to estimate the
contributions from sign-definite factors at a given value of the
boundary angle $z$. Maximizing the criterion, one could derive the
value $z^*$, which corresponds to the sign-definite observable
$\Delta\sigma_l(z^*)$. Since the observable is linear in the
coefficients $A$, $B$, and $C$, we introduce the following
criterion,
\begin{equation}
F=\frac{|\tilde{A}_{11}|+\omega_B |\tilde{B}_{11}| + \omega_C
|\tilde{C}_{1111}|}
 {\sum\limits_{{\rm all}~\tilde{A}}\left|\tilde{A}_{ij}\right|
  +\omega_B\sum\limits_{{\rm all}~\tilde{B}}\left|\tilde{B}_{ij}\right|
  +\omega_C\sum\limits_{{\rm all}~\tilde{C}}\left|\tilde{C}_{ijkn}\right|},
\end{equation}
where the positive `weights' $\omega_B\sim\zeta$ and
$\omega_C\sim\epsilon$ take into account the order of each term in
the inverse $Z'$ mass.

The numeric values of the `weights' $\omega_B$ and $\omega_C$ can
be taken from the present day bounds on the contact couplings
\cite{EWWG,LEP2FF02-03} or \cite{hep01}. As the computation shows,
the value of $z^*$ with the accuracy $10^{-3}$ depends on the
order of the `weight' magnitudes, only. So, in what follows we
take $\omega_B\sim .004$ and $\omega_C\sim 0.00004$.

The function $z^\ast(s)$ is the decreasing function of the
center-of-mass energy. It is tabulated for the LEP2 energies in
Tables 1-2. The corresponding values of the maximized function $F$
are within the interval $0.95<F<0.96$.

Since $\tilde{A}^l_{11}(s,z^*)<0$, $\tilde{B}^l_{11}(s,z^*)<0$ and
$\tilde{C}^l_{1111}(s,z^*)<0$, the observable
\begin{equation}\label{7}
 \Delta\sigma_l(z^*)=
 \left[\tilde{A}^l_{11}(s,z^*) +\zeta\tilde{B}^l_{11}(s,z^*)
 \right]\epsilon + \tilde{C}^l_{1111}(s,z^*)\epsilon^2
\end{equation}
is negative with the accuracy 4--5\%. Since this property follows
from the RG relations (\ref{2}) for the Abelian $Z'$ boson, the
observable $\Delta\sigma_l(z^*)$ selects the model-indepen\-dent
signal of this particle in the processes $e^+e^-\to l^+l^-$. It
allows to use the data on scattering into $\mu\mu$ and $\tau\tau$
pairs in order to estimate the Abelian $Z'$ coupling to the
axial-vector lepton currents.

\begin{table}
\caption{ The boundary angle and the observable for the scattering
into $\mu$ pairs at the one-loop level. }\label{tobsmu}\centering
\begin{center}
\begin{tabular}{|c|c|c|c|}\hline
 $\sqrt{s}$, GeV & $z^*$ & $F_{\rm max}$ & $\Delta\sigma_\mu(z^*)$ \\\hline\hline
 130 & 0.450 & 0.89 & $-729\epsilon-1792\epsilon\zeta-19636\epsilon^2$\\
 136 & 0.439 & 0.91 & $-709\epsilon-1859\epsilon\zeta-16880\epsilon^2$\\
 161 & 0.400 & 0.94 & $-643\epsilon-2183\epsilon\zeta-6890\epsilon^2$\\
 172 & 0.390 & 0.95 & $-619\epsilon-4099\epsilon\zeta-4099\epsilon^2$\\
 183 & 0.383 & 0.95 & $-599\epsilon-2545\epsilon\zeta-1334\epsilon^2$\\
 189 & 0.380 & 0.96 & $-586\epsilon-2635\epsilon\zeta-495\epsilon^2$\\
 192 & 0.380 & 0.96 & $-579\epsilon-2681\epsilon\zeta-63\epsilon^2$\\
 196 & 0.380 & 0.96 & $-571\epsilon-2745\epsilon\zeta-528\epsilon^2$\\
 200 & 0.378 & 0.95 & $-564\epsilon-2811\epsilon\zeta-1137\epsilon^2$\\
 202 & 0.376 & 0.96 & $-560\epsilon-2845\epsilon\zeta-1448\epsilon^2$\\
 205 & 0.374 & 0.96 & $-555\epsilon-2897\epsilon\zeta-1923\epsilon^2$\\
 207 & 0.372 & 0.96 & $-552\epsilon-2932\epsilon\zeta-2245\epsilon^2$\\
 \hline
\end{tabular}
\end{center}
\end{table}

\begin{table}
\caption{ The boundary angle and the observable for the scattering
into $\tau$ pairs at the one-loop level.
}\label{tobstau}\centering
\begin{center}
\begin{tabular}{|c|c|c|}\hline
 $\sqrt{s}$, GeV & $z^*$ &  $\Delta\sigma_\tau(z^*)$ \\\hline\hline
 130 & 0.460 & $-687\epsilon-1664\epsilon\zeta-25782\epsilon^2$\\
 136 & 0.442 & $-688\epsilon-1779\epsilon\zeta-20784\epsilon^2$\\
 161 & 0.400 & $-625\epsilon-2097\epsilon\zeta-10993\epsilon^2$\\
 172 & 0.391 & $-601\epsilon-2263\epsilon\zeta-8382\epsilon^2$\\
 183 & 0.385 & $-571\epsilon-2402\epsilon\zeta-7580\epsilon^2$\\
 189 & 0.380 & $-568\epsilon-2533\epsilon\zeta-5135\epsilon^2$\\
 192 & 0.380 & $-562\epsilon-2578\epsilon\zeta-4769\epsilon^2$\\
 196 & 0.379 & $-554\epsilon-2640\epsilon\zeta-4272\epsilon^2$\\
 200 & 0.378 & $-547\epsilon-2704\epsilon\zeta-3761\epsilon^2$\\
 202 & 0.377 & $-543\epsilon-2736\epsilon\zeta-3501\epsilon^2$\\
 205 & 0.374 & $-548\epsilon-2834\epsilon\zeta-1292\epsilon^2$\\
 207 & 0.372 & $-544\epsilon-2868\epsilon\zeta-1010\epsilon^2$\\
 \hline
\end{tabular}
\end{center}
\end{table}

Although the observable can be computed from the differential
cross-sections directly, it is also possible to recalculate it
from the total cross-sections and the forward-backward
asymmetries. The recalculation procedure has the proper
theoretical accuracy. Nevertheless, it allows to reduce the
experimental errors on the observable, since the published data on
the total cross-sections and the forward-backward asymmetries are
still more precise than the data on the differential
cross-sections.

The recalculation is based on the fact that the differential
cross-section can be approximated with a good accuracy by the
two-parametric polynomial in the cosine of the scattering angle
$z$:
\begin{equation}
\frac{d\sigma_l(s)}{dz} = \frac{d\sigma_l^{\rm SM}(s)}{dz} +
(1+z^2)\beta_l + z \eta_l +\delta_l(z),
\end{equation}
where $\delta_l(z)$ measures the difference between the exact and
the approximated cross-sections. The approximated cross-section
reproduces the exact one in the limit of the massless initial- and
final-state leptons and if one neglects the contributions of the
box diagrams.

Performing the angular integration, it is easy to obtain the
expression for the observable:
\begin{equation}
\Delta\sigma_l(z^*)=\sigma_l(z^*) - \sigma_l^{\rm SM}(z^*) =
(1-z^{*2})\eta_l -\frac{2\beta_l}{9} z^*(3+z^{*2})
+\tilde\delta_l(z^*),
\end{equation}
and for the total and the forward-backward cross-sections:
\begin{eqnarray}
\Delta\sigma^{\rm T}_l &=&\sigma^{\rm T}_l- \sigma_l^{\rm T,SM}
=\frac{8\beta_l}{9}
 +\tilde\delta_l(-1), \nonumber\\ \Delta\sigma^{\rm FB}_l &=&
\sigma^{\rm FB}_l-\sigma_l^{\rm FB,SM} =\eta_l +\tilde\delta_l(0).
\end{eqnarray}
Then, the factors $\beta_l$ and $\eta_l$ can be eliminated from
the observable:
\begin{eqnarray}
\Delta\sigma_l(z^*) &=& (1-z^{*2})\Delta\sigma^{\rm FB}_l
-\frac{3}{12} z^*(3+z^{*2})\Delta\sigma^{\rm T}_l +\xi_l.
\end{eqnarray}
The quantity $\xi_l$,
\begin{eqnarray}
\xi_l&=&\tilde\delta_l(z^*) -(1-z^{*2})\tilde\delta_l(0)
+\frac{3}{12} z^*(3+z^{*2})\tilde\delta_l(-1),
\end{eqnarray}
measures the theoretical accuracy of the approximation.

The forward-backward cross-section is related to the total one and
the forward-backward asymmetry by means of the following
expression
\begin{eqnarray}
\Delta\sigma^{\rm FB}_l &=& \Delta\sigma^{\rm T}_l \,\, A^{\rm
FB}_l +\sigma_l^{\rm T,SM} \,\, \Delta A^{\rm FB}_l.
\end{eqnarray}
As the computation shows, $\tilde\delta_l(z^*)\simeq 0.01
\Delta\sigma_{l}(z^*)$, $\tilde\delta_l(0)\simeq 0.007
\Delta\sigma^{\rm FB}_l$, and $\tilde\delta_l(-1)\simeq -0.07
\Delta\sigma^{\rm T}_l$ at the LEP2 energies. Taking into account
the experimental values of the total cross-sections and the
forward-backward asymmetries at the LEP2 energies
($\Delta\sigma^{\rm T}_l\simeq 0.1$pb, $\sigma_l^{\rm T,SM}\simeq
2.7$pb, $\Delta A_l^{\rm FB}\simeq 0.04$, $A_l^{\rm FB}\simeq
0.5$), one can estimate the theoretical error as $\xi_l\simeq
0.003{\rm pb}$. At the same time, the corresponding statistical
uncertainties on the observable are larger than 0.06pb. Thus, the
proposed approximation is quite good and can be successfully used
to obtain more accurate experimental values of the observable.

\section{Data fit and Conclusions}

To search for the model-independent signals of the Abelian
$Z'$-boson we will analyze the introduced observable
$\Delta\sigma_l (z^\ast)$ on the base of the LEP2 data set. In the
lower order in $m^{-2}_{Z'}$ the observable (\ref{7}) depends on
one flavor-independent parameter $\epsilon$,
\begin{equation}
 \Delta\sigma^{\rm th}_l(z^*)=
 \tilde{A}^l_{11}(s,z^*)\epsilon + \tilde{C}^l_{1111}(s,z^*)\epsilon^2,
\end{equation}
which can be fitted from the experimental values of
$\Delta\sigma_\mu (z^\ast)$ and $\Delta\sigma_\tau (z^\ast)$. As
we noted above, the sign of the fitted parameter ($\epsilon >0$)
is a characteristic feature of the Abelian $Z'$ signal.

In what follows we will apply the usual fit method based on the
likelihood function. The central value of $\epsilon$ is obtained
by the minimization of the $\chi^2$-function:
\begin{equation}
\chi^2(\epsilon) = \sum_{n} \frac{\left[\Delta\sigma^{\rm
ex}_{\mu,n}(z^*)- \Delta\sigma^{\rm th}_\mu(z^*)\right]^2}
{\delta\sigma^{\rm ex}_{\mu,n}(z^*)^2},
\end{equation}
where the sum runs over the experimental points entering a data
set chosen. The $1\sigma$ confidence-level interval $(b_1,b_2)$
for the fitted parameter is derived by means of the likelihood
function ${\cal L}(\epsilon)\propto\exp[-\chi^2(\epsilon)/2]$. It
is determined by the equations:
\begin{equation}
\int\nolimits_{b_1}^{b_2}{\cal L}(\epsilon ')d\epsilon ' = 0.68,
 \quad
{\cal L}(b_1)={\cal L}(b_2).
\end{equation}

To compare our results with those of Refs. \cite{EWWG,LEP2FF02-03}
we introduce the contact interaction scale
\begin{equation}
\Lambda^2 = 4m^2_Z\epsilon^{-1}.
\end{equation}
This normalization of contact couplings is admitted in Refs.
\cite{EWWG,LEP2FF02-03}. We use again the likelihood method to
determine a one-sided lower limit on the scale $\Lambda$ at the
95\% confidence level. It is derived by the integration of the
likelihood function over the physically allowed region
$\epsilon>0$. The exact definition is
\begin{equation}
\Lambda=2m_Z (\epsilon^*)^{-1/2}, \quad \int_{0}^{\epsilon^*}{\cal
L}(\epsilon ')d\epsilon ' = 0.95\int_{0}^{\infty}{\cal L}(\epsilon
')d\epsilon '.
\end{equation}

We also introduce the probability of the Abelian $Z'$ signal as
the integral of the likelihood function over the positive values
of $\epsilon$:
\begin{equation}
P=\int\nolimits_{0}^{\infty} L(\epsilon ')d\epsilon '.
\end{equation}

Actually, the fitted value of the contact coupling $\epsilon$
originates mainly from the leading-order term in the inverse $Z'$
mass in Eq. (\ref{7}). The analysis of the higher-order terms
allows to estimate the constraints on the $Z'$ mass alone.
Substituting $\epsilon$ in the observable (\ref{7}) by its fitted
central value, $\bar{\epsilon}$, one obtains the expression
\begin{equation}
 \Delta\sigma_l(z^*)=
 \left[\tilde{A}^l_{11}(s,z^*)
  +\zeta \tilde{B}^l_{11}(s,z^*) \right]\bar\epsilon +
 \tilde{C}^l_{1111}(s,z^*)\bar\epsilon^2,
\end{equation}
which depends on the parameter $\zeta=m^2_Z/m^2_{Z'}$. Then, the
central value on this parameter and the corresponding 1$\sigma$
confidence level interval are derived in the same way as those for
$\epsilon$.

To fit the parameters $\epsilon$ and $\zeta$ we start with the
LEP2 data on the total cross-sections and the forward-backward
asymmetries \cite{EWWG,LEP2FF02-03}. Those data are converted into
the experimental values of the observable $\Delta\sigma_l(z^\ast)$
with the corresponding errors $\delta\sigma_l(z^\ast)$ by means of
the following relations:
\begin{eqnarray}
 \Delta\sigma_l(z^\ast)
 &=&
 \left[
 A_l^{\rm FB}\left(1-z^{\ast 2}\right)
 -\frac{z^\ast}{4}\left(3 +z^{\ast 2}\right)
 \right] \Delta\sigma_l^{\rm T}
 \nonumber\\&&
 + \left(1 - z^{\ast 2}\right)
  \sigma_l^{\rm T,SM} \Delta A_l^{\rm FB},
 \nonumber\\
 \delta\sigma_l(z^\ast)^2
 &=&
 {\left[
 A_l^{\rm FB}\left(1-z^{\ast 2}\right)
 -\frac{z^\ast}{4}\left(3 +z^{\ast 2}\right)
 \right]}^2 (\delta\sigma_l^{\rm T})^2
 \nonumber\\&&
 +{\left[
 \left(1 - z^{\ast 2}\right)
 \sigma_{l}^{\rm T,SM}
 \right]}^2 (\delta A_l^{\rm FB})^2.
\end{eqnarray}

We perform the fits assuming several data sets, including the
$\mu\mu$, $\tau\tau$, and the complete $\mu\mu$ and $\tau\tau$
data, respectively. The results are presented in Table
\ref{tfitrecalc}.
\begin{table}
\caption{ The contact coupling $\epsilon$ with the 68\%
confidence-level uncertainty, the 95\% confidence-level lower
limit on the scale $\Lambda$, the probability of the $Z'$ signal,
$P$, and the value of $\zeta=m^2_Z/m^2_{Z'}$ as a result of the
fit of the observable recalculated from the total cross-sections
and the forward-backward asymmetries.
}\label{tfitrecalc}\centering
\begin{tabular}{l|c|c|c|c}
\hline\hline Data set & $\epsilon$ & $\Lambda$, TeV & $P$ &
$\zeta$
 \\ \hline
 \multicolumn{5}{c}{Winter 2002} \\ \hline
 $\mu\mu$ & $0.0000482^{+0.0000496}_{-0.0000493}$
 & 15.7
 & 0.83
 & $0.007\pm 0.215$
 \\
 $\tau\tau$ & $0.0000016^{+0.0000661}_{-0.0000656}$
 & 16.0
 & 0.51
 & $-0.052\pm 8.463$
 \\
 $\mu\mu$ and $\tau\tau$ & $0.0000313^{+0.0000396}_{-0.0000395}$
 & 18.1
 & 0.78
 & $0.006\pm 0.264$
 \\ \hline
 \multicolumn{5}{c}{Summer 2002} \\ \hline
 $\mu\mu$ & $0.0000366^{+0.0000489}_{-0.0000486}$
 & 16.4
 & 0.77
 & $0.009\pm 0.278$
 \\
 $\tau\tau$ & $-0.0000266^{+0.0000643}_{-0.0000639}$
 & 17.4
 & 0.34
 & $-0.001\pm 0.501$
 \\
 $\mu\mu$ and $\tau\tau$ & $0.0000133^{+0.0000389}_{-0.0000387}$
 & 19.7
 & 0.63
 & $0.017\pm 0.609$
 \\ \hline\hline
\end{tabular}
\end{table}
As is seen, the more precise $\mu\mu$ data demonstrate the signal
of about 1$\sigma$ level. It corresponds to the Abelian $Z'$-boson
with the mass of order 1.2--1.5TeV if one assumes the value of
$\tilde\alpha=\tilde{g}^2/4\pi$ to be in the interval 0.01--0.02.
No signal is found by the analysis of the $\tau\tau$
cross-sections. The combined fit of the $\mu\mu$ and $\tau\tau$
data leads to the signal below the 1$\sigma$ confidence level.

Note that the mean values of $\epsilon$ have changed by 20\% in
comparison with the Winter 2002 data, whereas the uncertainties
remain approximately the same. So, the Abelian $Z'$ signal will be
probably picked up at no more than 1$\sigma$ confidence level when
the final LEP2 data on $e^+e^-\to\mu^+\mu^-,\tau^+\tau^-$ will be
completed.

Being governed by the next-to-leading contributions in
$m^{-2}_{Z'}$, the fitted values of $\zeta$ are characterized by
significant errors. The $\mu\mu$ data set gives the central value
which corresponds to $m_{Z'}\simeq 1.1$ TeV.

We also perform a separate fit of the parameters based on the
direct calculation of the observable from the differential
cross-sections. The complete set of the available data
\cite{diffcs} is used (see Table \ref{tdsset}).
\begin{table}\label{tdsset}
\caption{The differential cross-sections used for fitting.
 The letters F and P mark the final and the preliminary data,
 respectively.}
\begin{center}
\begin{tabular}{|c|c|c|c|c|}
\hline\hline $\sqrt{s}$, GeV & ALEPH & DELPHI & L3 & OPAL
 \\\hline 
 130 &   &   &   & F \\ 
 136 &   &   &   & F \\ 
 161 &   &   &   & F \\ 
 172 &   &   &   & F \\ 
 183 &   & F & F & F \\ 
 189 & P & F & F & F \\ 
 192 & P & P &   & P \\ 
 196 & P & P &   & P \\ 
 200 & P & P &   & P \\ 
 202 & P & P &   & P \\ 
 205 & P & P &   & P \\ 
 207 & P & P &   & P \\\hline\hline
\end{tabular}
\end{center}
\end{table}
The results are given in Table \ref{tfitds}.
\begin{table}
\caption{ The contact coupling $\epsilon$ with the 68\%
confidence-level uncertainties, computed from the differential
cross-sections. }\label{tfitds}\centering
\begin{center}
\begin{tabular}{|l|r|r|r|}
\hline 
 & $\mu$ data
 & $\tau$ data
 & $\mu$ and $\tau$ data \\\hline
 \rule{0cm}{0.4cm} ALEPH
 & $0.00014^{+0.00068}_{-0.00069}$
 & $-0.00007^{+0.00120}_{-0.00120}$
 & $0.00009^{+0.00059}_{-0.00060}$ \\ 
 \rule{0cm}{0.4cm} DELPHI
 & $-0.00010^{+0.00070}_{-0.00070}$
 & $0.00000^{+0.00140}_{-0.00140}$
 & $-0.00008^{+0.00062}_{-0.00063}$ \\ 
 \rule{0cm}{0.4cm} L3
 & $0.00013^{+0.00043}_{-0.00043}$
 & $0.00024^{+0.00053}_{-0.00054}$
 & $0.00018^{+0.00033}_{-0.00033}$ \\ 
 \rule{0cm}{0.4cm} OPAL
 & $0.00028^{+0.00074}_{-0.00075}$
 & $-0.00017^{+0.00120}_{-0.00120}$
 & $0.00015^{+0.00063}_{-0.00063}$ \\\hline 
 \rule{0cm}{0.4cm} Combined
 & $0.00012^{+0.00028}_{-0.00030}$
 & $0.00012^{+0.00043}_{-0.00043}$
 & $0.00012^{+0.00024}_{-0.00024}$
\\\hline
\end{tabular}
\end{center}
\end{table}
As is seen, the experimental uncertainties of the data on the
differential cross-sections are of one order larger than the
corresponding errors of the total cross-sections and the
forward-backward asymmetries. These data also provide the larger
values of the contact coupling $\epsilon$. As for the more precise
$\mu\mu$ data, three of the LEP2 Collaborations demonstrate
positive values of $\epsilon$. The combined $\epsilon$ is also
positive and remains practically unchanged by the incorporation of
the $\tau\tau$ data.

Now, we compare the fits based on the differential cross-sections
and the total cross-sections. As it was mentioned in the previous
section, the indirect computation of the observable from the total
cross-sections and the forward-backward asymmetries inspires some
insufficient theoretical uncertainty about 2\% of the statistical
one. It also increases the statistical error because of the
recalculation procedure. Nevertheless, the uncertainty of the
fitted parameter $\epsilon$ within the recalculation scheme is of
one order less than that for the direct computation from the
differential cross-sections. This difference is explained by the
different accuracy of the available experimental data on the
differential and the total cross-sections. If the final LEP2
differential cross-sections will be so accurate as the present
data on the total cross-sections, the direct computation of the
observable can reduce, in principle, the uncertainty of the fitted
coupling $\epsilon$.

To compare our results with the fits of the contact couplings
presented by the LEP Collaborations in Refs.
\cite{EWWG,LEP2FF02-03} let us briefly describe the approach used
therein. Since only one parameter of new physics can be
successfully fitted, the LEP Collaborations usually discuss eight
`models' (LL, RR, LR, RL, VV, AA, A0, V0) which assume specific
helicity couplings between the initial-state and the final-state
fermion currents. Each model is described by only one non-zero
four-fermion coupling, while others are set to zero. For example,
in the LL model the non-zero coupling of left-handed fermions is
taken into account. The signal of new physics is fitted by
considering the interference of the SM amplitude with the contact
four-fermion term. Whatever physics beyond the SM exists, it has
to manifest itself in some contact coupling mentioned. Hence, it
is possible to find the low limit on the masses of the states
responsible for the interactions considered. In principle, a
number of states may contribute into each of the models.
Therefore, the purpose of the fit described by these models is to
find any signal of new physics. No specific types of new particles
are considered in this analysis.

As it was shown, the characteristic signal of the Abelian $Z'$
boson is related to the flavor-independent couplings to the
axial-vector currents. To pick up the signal we construct the
observable which is dominated by the axial-vector couplings. The
contributions of the rest couplings are suppressed in the
observable by the special choice of the kinematic parameters. In
this regard, let us turn to the helicity `models' of Refs.
\cite{EWWG,LEP2FF02-03} and compare our results with the fit for
the axial model (AA). As it follows from the present analysis,
this model could be sensitive to the signals of the Abelian $Z'$
boson. Of course, the parameters $\epsilon$ in Refs.
\cite{EWWG,LEP2FF02-03} (in what follows we will mark it as
$\epsilon_{\rm EWWG}$) and $\epsilon$ in the present paper are not
the same quantity. First, they are normalized by different factors
and related as $\epsilon_{\rm EWWG}=-\epsilon m^{-2}_Z/4$. Second,
as we already noted, in the AA model the $Z'$ couplings to the
vector fermion currents are set to zero, therefore it is able to
describe only some particular case of the Abelian $Z'$ boson.
Moreover, in this model both the positive and the negative values
of $\epsilon_{\rm EWWG}$ are considered, whereas in our approach
only the positive $\epsilon$ values (which correspond to the
negative $\epsilon_{\rm EWWG}$) are permissible. As the value of
the four-fermion contact coupling in the AA model is dependent on
the lepton flavor, the Abelian $Z'$ induces the axial-vector
coupling which is universal for all lepton types. Considering the
Winter 2002 data \cite{EWWG}, it is interesting to note that the
fitted value of $\epsilon_{\rm EWWG}$ in the AA model for the
$\mu^+\mu^-$ final states ($-0.0025^{+0.0018}_{-0.0023}\mbox{
TeV}^{-2}$) as well as the value derived under the assumption of
the lepton universality ($-0.0018^{+0.0016}_{-0.0019}\mbox{
TeV}^{-2}$) are similar to our results which correspond to
$-0.0015\pm 0.0015\mbox{ TeV}^{-2}$ and $-0.0009^\pm 0.0012\mbox{
TeV}^{-2}$, respectively. As for the Summer 2002 data
\cite{LEP2FF02-03}, the value of $\epsilon_{\rm EWWG}$ under the
assumption of the lepton universality is available only
($-0.0013^\pm 0.0017\mbox{ TeV}^{-2}$). It is close to our result
for the $\mu\mu$ process ($-0.0011\pm 0.0015\mbox{ TeV}^{-2}$).
However, the central value of $\epsilon_{\rm EWWG}$ is about three
times greater than the corresponding one for the combined $\mu$
and $\tau$ data ($-0.0004^\pm 0.0012\mbox{ TeV}^{-2}$). Thus, the
signs of the central values in the AA model agree with our
results, and the uncertainties are of the same order. The fitted
values of the 95\% confidence-level lower limit on the scale
$\Lambda$ are again in a good accordance with the corresponding
values of $\Lambda^-$ for the AA model of \cite{LEP2FF02-03}. So,
we come to a conclusion that the AA model is mainly responsible
for signals of the Abelian $Z'$ gauge boson although a lot of
details concerning its interactions is not accounted for within
this fit.

It worth to mention the recent paper by Chivukula and Simmons
\cite{fczp} who derived model-dependent bounds on the mass of the
$Z'$ boson for flavor-changing technicolor models. It has been
obtained that in these models $m_{Z'}$ is heavier than about 1
TeV. It is interesting to note that this value is very close to
our model-independent result which corresponds to the
flavor-conserved case.

As it follows from the present analysis, the Abelian $Z'$ boson
has to be light enough to be discovered at the LHC. On the other
hand, the LEP2 data on the processes
$e^+e^-\to\mu^+\mu^-,\tau^+\tau^-$ do not provide the necessary
statistics for the detection of the model-independent signal of
the Abelian $Z'$ boson at more than 1$\sigma$ confidence level.
So, it is of interest to find the observables for other scattering
processes in order to increase the data set. In this regard, let
us note the paper \cite{bourilkov} where the helicity `models'
were applied to the Bhabha scattering $e^+e^-\to e^+e^-(\gamma)$.
As it was shown therein, the AA model demonstrates the
$2\sigma$--level deviation from the SM. However, these deviations
could not be interpreted directly as the signal of the Abelian
$Z'$ boson because of the reasons mentioned above. Therefore, it
seems to us perspective to find the observable for the Abelian
$Z'$ search in the process $e^+e^-\to e^+e^-$.

\end{document}